\documentclass{article}
\usepackage{spconf,amsmath,graphicx,hyperref,times}
\usepackage{amssymb, mathtools}
\usepackage{multirow, booktabs, hhline, array}
\usepackage{makecell}
\usepackage{xcolor}

\def\thline{\noalign{\hrule height 1.0pt}}

\title{Unsupervised multi-channel separation and adaptation}

\name{Cong Han$^{1*}$\thanks{$^*$Work performed during internship at Google.}, Kevin Wilson$^{2}$, Scott Wisdom$^{2}$, John R.\ Hershey$^{2}$}
\address{$^{1}$Columbia University, $^{2}$Google }

\begin{document}
\ninept
\maketitle
\begin{abstract}
\end{abstract}

A key challenge in machine learning is to generalize from training data to an application domain of interest.  This work extends the recently-proposed mixture invariant  training (MixIT) algorithm to perform unsupervised learning in the multi-channel setting.  We use MixIT to train a model on far-field microphone array recordings of  overlapping  reverberant  and  noisy speech from the AMI Corpus. The models are trained on both supervised and unsupervised training data, and are tested on real AMI recordings containing overlapping speech. To objectively evaluate our models, we also use a synthetic multi-channel AMI test set. Holding  network  architectures  constant,  we  find that semi-supervised fine-tuning of a model pretrained on a large and diverse single-channel dataset yields the largest improvement to SI-SNR and to human listening ratings across synthetic and real datasets, outperforming supervised models trained on well-matched synthetic data.  Our results demonstrate that unsupervised learning through MixIT enables model adaptation on both single- and multi-channel real-world speech recordings.

\begin{keywords}
multi-channel, speech separation
\end{keywords}
\section{Introduction}
\label{sec:intro}

Separation and enhancement of sounds is an important and challenging problem in machine perception. Recent progress has been made in multi-channel speech separation and enhancement using neural network methods. Neural network-based mask estimators \cite{Wang2018, gu2019end, han2020real} and neural beamformers \cite{heymann2015blstm,erdogan2016improved,ochiai2020beam, wang2021sequential,sainath2017multichannel,luo2019fasnet, zhang2021adl} require supervised training data with input sound mixtures paired with isolated sounds as ground-truth targets. However, it is not feasible to record such pairs of isolated sounds and their mixtures in a real environment.  Thus for supervised training, input mixtures are constructed by synthetically mixing isolated recordings of the target sources. Unfortunately, this can result in a mismatch in the distribution of sound types and acoustic conditions between the simulated sound mixtures and real-world audio. For example, conversational speech is mismatched to the read speech that is typically used to train speech enhancement and separation models. Additionally, simulating multi-channel acoustics that match real-world scenarios is challenging due to potential source motion, varying microphone array geometries and microphone directivities, and other factors.

Unsupervised methods have helped to overcome the mismatch problem by directly training on real recordings from the target domain, without the need for ground-truth isolated sources. One category of approaches uses spatial information to first cluster sound sources in space in an unsupervised manner. These cluster labels are then employed as pseudo-targets for supervised training of neural networks \cite{tzinis2019unsupervised, drude2019unsupervised-dpcl, togami2020unsupervised, han2022multi}. \cite{saijo22b_interspeech} introduces a spatial loss that forces the beamforming vector estimated for a target source to be aligned with the corresponding steering vector while being orthogonal for interfering sources. However, this category of approaches relies on clustering or localization algorithms, such as k-means, Gaussian mixture model \cite{ito2016complex}, or multiple signal classification (MUSIC) \cite{schmidt1986multiple}, which are based on simple statistical models and cannot be trained for accuracy. In the case of co-located sources and strongly reverberant environments, where these algorithms may make errors, the separation performance usually suffers.

Mixture invariant training (MixIT) \cite{wisdom2020mixit} is a recent unsupervised approach that has demonstrated competitive single-channel sound separation performance. MixIT uses mixtures of mixtures as the ``noisy'' input and uses the individual mixtures as weak references. The model estimates individual sound sources that can be recombined to reconstruct the original reference mixtures. Note that MixIT incurs another type of mismatch in which there are more active sources in the mixture-of-mixtures than there are in an individual mixture.  Experiments have shown that when unsupervised training with MixIT and supervised training are performed jointly, the mismatch introduced by one training method is mitigated by the other. MixIT has been shown to be effective at adapting  single-channel speech separation models to real-world meetings \cite{Sivaraman2022}. 
 
In this work, we extend MixIT to multi-channel data, allowing the model to use both spatial and spectral information to better separate sound sources. We use a separation model with multi-channel input and multi-channel output that employs a temporal convolutional network (TCN) \cite{luo2019conv, kavalerov2019universal} and a transform-average-concatenate (TAC) module \cite{luo2020end, yoshioka2022vararray}, which enables the model to be applied to any number of microphones and any array geometries. This flexibility is particularly advantageous for models trained on diverse real-world meeting data captured by different microphone arrays. We show that when used with our flexible multi-microphone neural network, MixIT training on real mixtures improves separation and enhancement of speech on real meetings containing spontaneous speech and recorded with multiple microphones.  Better separation of such scenarios has many practical uses such as better automatic meeting transcripts and better telephony experiences.%

\begin{figure*}[!t]
    \centering
    \includegraphics[width=2.1\columnwidth]{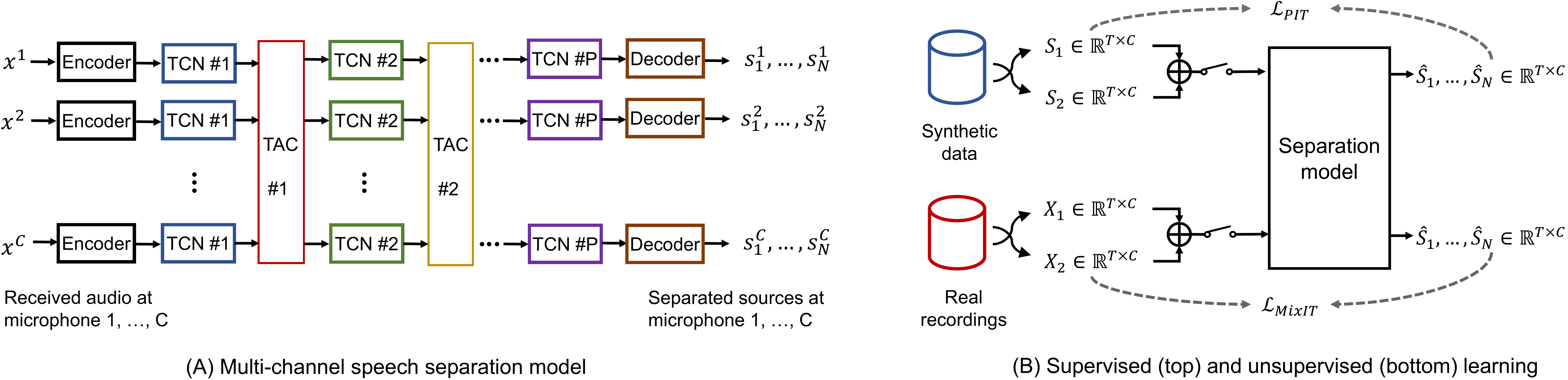}
    \caption{(A) The architecture of the proposed multi-channel input and multi-channel output speech separation model. Blocks with the same color share parameters. (B) The schematic of supervised learning with PIT on synthetic data (top) and unsupervised learning with MixIT on real recordings (bottom). }
    \label{fig:diag}
    \vspace{-5pt}
\end{figure*}

\section{Methods}
\label{sec:Methods}

Fig.~\ref{fig:diag}(A) describes our multi-channel speech separation model that accepts waveform input from multiple microphones and produces the multi-channel image of each source.  This separation model can be trained through supervised learning with permutation invariant training (PIT) \cite{yu2017permutation}, through unsupervised learning with multi-channel mixture invariant training (MC-MixIT), or through a combination of both as shown in Fig.~\ref{fig:diag}(B).  We describe the extension of MixIT to MC-MixIT in Section \ref{sec:mc-mixit}.

\subsection{Multi-channel speech separation model}
\label{sec:baseline-model}
The model is a variant of a single-channel separation model, TDCN++ \cite{kavalerov2019universal}.  To enable use on multi-channel audio, we interleave transform-average-concatenate (TAC) layers \cite{luo2020end} between temporal convolutional neural networks (TCNs) \cite{luo2019conv} to exploit spatio-temporal information across channels. The model shares some similarities with VarArray \cite{yoshioka2022vararray}. Both models are designed to be invariant to microphone array geometry and the number of microphones used. However, there are two main differences between our model and VarArray. First, VarArray calculates a feature set from STFT coefficients, while our model takes the raw waveform directly as input. Second, VarArray merges all channels at an intermediate layer and estimates a single time-frequency mask for each source, while our model estimates a multi-mic waveform for each source.

Given a C-channel time-domain signal $\boldsymbol{X} \in \mathbb{R}^{T \times C}$, where $T$ is the duration of the signal, we apply a linear encoder followed by a ReLU activation to transform each channel $\boldsymbol{x}^c \in \mathbb{R}^{T}, c=1, ..., C$ to a two-dimensional representation $\boldsymbol{E}^c \in \mathbb{R}^{F \times L}$, where F is the number of encoder bases and $L$ is the number of time frames. Then, $\left\{\boldsymbol{E}^c\right\}_{c=1}^C$ is fed into a series of alternating TCN and TAC layers. A TCN block comprises multiple dilated convolution layers, with the output of the $i^{\mathrm{th}}$ TCN block in channel $c$ denoted as $\boldsymbol{P}_i^c \in \mathbb{R}^{K \times L}$, where K is the number of features. To extract cross-channel features, we employ TAC layers that aggregate the outputs from each channel, extract cross-channel information, and feed it back to individual channels. Following the approach in \cite{yoshioka2022vararray}, the output of the i-th TAC layer in channel c, denoted as $\boldsymbol{Q}_i^c \in \mathbb{R}^{2K \times L}$, is:

\begin{align}
    \boldsymbol{Q}_i^c = \left[ \text{ReLU}(\boldsymbol{W}_i\boldsymbol{P}_i^c), \frac{1}{C}\sum_c{\text{ReLU}(\boldsymbol{U}_i\boldsymbol{P}_i^c)}\right],
\end{align}
where $\boldsymbol{W}_i, \boldsymbol{U}_i \in \mathbb{R}^{K \times K}$ are linear transforms. After the final TCN block, we use a sigmoid activation to predict a mask for each source in each channel and use a linear decoder to transform the masked representation back to the waveform. Note that if we remove all the TAC layers, the architecture is equivalent to independently applying a single-channel TDCN++ to individual channels. The TCN blocks process each channel locally while the TAC layers allow for inter-channel information flow.

\subsection{MixIT}
\label{sec:mixit}
MixIT uses $N$ (typically $N=2$) reference mixtures $\boldsymbol{x}_n \in \mathbb{R}^{T}$, which are the columns of a matrix $\boldsymbol{X} \in \mathbb{R}^{T \times N}$. A mixture of mixtures (MoM) is formed by summing these reference mixtures to produce $\boldsymbol{\hat{x}}=\sum_N{x_n}$. The network then generates $M>N$ estimated sources $\left\{\boldsymbol{s}_m \in \mathbb{R}^{T}\right\}_{m=1}^M$, which are the columns of a matrix $\boldsymbol{S} \in \mathbb{R}^{T \times M}$. The MixIT loss estimates a mixing matrix $\boldsymbol{A} \in \mathbb{B}$, where $\mathbb{B}=\left\{0,1\right\}^{M\times N}$ is a constrained set of $M \times N$ binary matrices where each row sums to 1: that is, the set of matrices which assign each estimated source $\boldsymbol{s}_m$ to one of the reference mixtures $\boldsymbol{x}_n$. Given the mixing matrix, a signal level loss, $\mathcal{L}$, measures the error between reference mixtures and their assigned estimates:
\begin{align}
    \mathcal{L}_{\text{MixIT}}(\boldsymbol{X}, \boldsymbol{S}) = \min_{A \in \mathbb{B}}\mathcal{L}(\boldsymbol{X}, \boldsymbol{S}\boldsymbol{A}).
\end{align}
where $\mathcal{L}$ typically operates column-wise, so that $\mathcal{L}(\boldsymbol{X}, \boldsymbol{S}) = \sum_{n}{\mathcal{L}(\boldsymbol{x}_n, (\boldsymbol{S}\boldsymbol{A})_n)}$. In this paper, $\mathcal{L}$ is negative thresholded SNR:
\begin{align}
\label{eqn:snr}
    \mathcal{L}(\boldsymbol{y}, \hat{\boldsymbol{y}}) = 10\,\text{log}_{10}\left(\frac{||\boldsymbol{y}||^2}{||\hat{\boldsymbol{y}} - \boldsymbol{y}||^2 + \tau||\boldsymbol{y}||^2}\right),
\end{align}
where $\tau$ is a soft limit on the maximum SNR. We select $\tau = 0.001$.

\subsection{Multi-Channel MixIT}
\label{sec:mc-mixit}
We extend MixIT by applying the same mixing matrix to all the channels of each source. We write the multi-channel references $\boldsymbol{X}^c$ and sources $\boldsymbol{S}^c$ at channel $c$. The MixIT loss requires finding the optimal mixing matrix $\boldsymbol{A} \in \mathbb{B}$ across all channels:
\begin{align}
    \mathcal{L}_{\text{MC-MixIT}}(\left\{\boldsymbol{X}^c\right\}, \left\{\boldsymbol{S}^c\right\}) = \min_{A \in \mathbb{B}}\sum_c\mathcal{L}(\boldsymbol{X}^c, \boldsymbol{S}^c\boldsymbol{A}).
\end{align}
Sharing $\boldsymbol{A}$ across microphones encourages the order of the separated sources in the model outputs to be consistent across all channels.  

\section{Experiments}
\label{sec:experiments}

Our experimental approach largely follows that of \cite{Sivaraman2022}. We conducted experiments using the AMI Corpus \cite{CarlettaJ2006AMI} of meeting room recordings for evaluation data and as one source of training data.  Hyperparameters were fixed to values in Table \ref{tab:hparams} across all experiments; only training procedures were varied.

\begin{table}[!ht]
\small
\centering
\caption{Hyperparameter values.}
\label{tab:hparams}

\begin{tabular}{c|c|c} \hline
Category & Hyperparameter & value \\ \hline
\multirow{9}{*}{Model} 
& TCN superblocks & 4 \\
& TCN blocks per superblock & 8 \\
& TCN kernel width & 3 \\
& TCN window size & 64 samples \\
& TCN hop size & 32 samples \\
& TCN bottleneck dim & 128 \\
& TCN conv channels & 512 \\
& TAC projection dim & 128 \\
& \# of output sources & 8 \\\hline
\multirow{3}{*}{Data} 
& Unsup example len & 10 seconds \\
& Sup example len & 5 seconds \\
& Audio sample rate & 16 kHz \\\hline
\multirow{5}{*}{Training} 
& Trainable weights & 4.7 million \\
& Optimizer & Adam \\
& Batch size & 256 \\
& Learning rate & $3 * 10^{-4}$ \\ 
& Training steps & 1 million \\
\hline \end{tabular}
\end{table}

\subsection{Training}
\label{sec:training}

All models have $M=8$ output sources, though depending on the training configuration, there may be fewer target sources.  When there are fewer than 8 target sources, our negative thresholded SNR loss is applied to only the non-zero sources, and we rely on mixture consistency \cite{wisdom2018consistency} to push unused outputs to zero.

Our training configurations are shown in Figure \ref{fig:diag} (B).  For unsupervised (MixIT) training, we use mixtures of randomly chosen segments from the AMI Corpus.  Our AMI training split is 71 hours. %

For supervised (PIT) training, we use single-talker segments of the AMI Corpus to synthesize training mixtures where each source has only a single active talker, as identified using the AMI 
annotations.
To generate each training example, two such segments are taken from the same room but from different speakers, who we refer to as speaker 1 and speaker 2.  We wish to use these as references that are added together to create a synthetic input mixture.  However, these single-speaker clips contain some background noise.  We experiment with two different approaches for dealing with the background noise when constructing references.  

The first approach addresses this problem by creating a cleaner reference following the procedure described in \cite{Sivaraman2022} to create ``synthetic overlapping AMI'' (referred to as ``synth AMI'' in Table \ref{tab:crosseval} and Table \ref{tab:ami}). To define a nearly noise-free speech reference, we use the headset mic recording for speaker 2's segment and find the multi-channel filter that optimally matches the microphone array signal.  We define that multi-channel filtered headset signal to be the training target for speaker 2.  We define the training target for speaker 1 to be the microphone array recording of speaker 1's segment (including background noise), and we define a third training target that is the residual from speaker 2's microphone array recording after subtracting the filtered headset target.  These three targets add up to the sum of the two microphone array segments, but they are asymmetric with respect to the speakers.  Because the speaker 2 reference is cleaner, we focus on speaker 2 during evaluation, below.  See \cite{Sivaraman2022} for additional details.

In the second approach, we directly use the array signals for speaker 1 and speaker 2 as references; we refer to this as ``mixed AMI'' in Table \ref{tab:ami}.  A caveat with this approach is that both reference signals contain some background noise, and thus the mixture contains double background noise.  As training targets, the noisy references may lead the model to preserve noise in its speech estimates.  For evaluation these references may not be as accurate as desired.

For 1-microphone training, we use only the first channel of the AMI recordings. For 2-microphone, 4-microphone, and 8-microphone training, we use 2, 4, and 8 microphones from the circular table-mounted microphone array at 180$^{\circ}$, 90$^{\circ}$, and 45$^{\circ}$ separation from one another, respectively. For all model evaluation, we evaluate only the first channel of the output.  Train, validation, and test splits follow the standard AMI ``full-corpus'' partition of meetings.

As a baseline and as a model from which to warm-start, we train a single-channel model on 1600 hours of audio from videos in the YFCC100M corpus
with train, validation, and test splits from \cite{wisdom2020mixit}.

All models are trained with a mixture consistency hard constraint \cite{wisdom2018consistency} on their outputs and with feature-wise layer normalization as described in \cite{kavalerov2019universal}.  We train all models for one million steps.

\subsection{Evaluation}
\label{sec:eval}

To objectively evaluate our methods on synthetic AMI, we use scale-invariant signal-to-noise ratio improvement (SI-SNRi) with the filtered headset signal as reference \cite{LeRoux2018a}. To measure subjective audio quality, we use the multiple stimulus with hidden reference and anchors (MUSHRA) \cite{itu2014mushra} for the synthetic AMI evaluation dataset, with the filtered headset as reference. For real AMI data, where we do not have a known clean reference, we adopt a variant of MUSHRA that allows for imperfect references called MUSHIRA \cite{Sivaraman2022}. For MUSHIRA on real AMI data, the imperfect reference is a headset recording of a target speaker that contains cross-talk. For all listening tests, audio is presented diotically, with the signal corresponding to the first microphone being presented to both ears.  We collect 5 ratings per example for both MUSHRA and MUSHIRA.

We trained two instances of each configuration and report averages across model instances for all metrics in Tables \ref{tab:crosseval} and \ref{tab:ami}.

\begin{table}[!t]
    \small
	\centering
	\caption{Cross-evaluation by number of mics. Values are SI-SNRi in dB.  ``S1'' and ``S2'' refer to the full-duration speaker and the overlapping speaker, respectively.  Due to space constraints, we report results on only ``synth AMI'' training data without warm start.}
	\scalebox{0.8}{
    \begin{tabular}{c|c|cc|cc|cc|cc}
    \thline
    \# of mics  & 
    Training  & \multicolumn{2}{c|}{1-mic eval} & \multicolumn{2}{c|}{2-mic eval} & \multicolumn{2}{c|}{4-mic eval} & \multicolumn{2}{c}{8-mic eval} \\
    - training & method & S1 & S2 & S1 & S2 & S1 & S2 & S1 & S2 \\
    \hline
    \multirow{3}{*}{1-mic} & Sup & 4.3 & 6.0 & 4.5 & 6.5 & 4.5 & 6.5 & 4.5 & 6.6   \\
    & Unsup & 3.7 & 10.0 & 3.6 & 9.9 & 3.6 & 9.9 & 3.6 & 9.9 \\
    & Semi & \bf{6.0} & \bf{10.3} & 6.1 & 10.6 & 6.1 & 10.6 & 6.1 & 10.6 \\
    \hline
    \multirow{3}{*}{2-mic} & Sup & 1.8 & -3.5 & 5.8 & 7.9 & 6.1 & 8.4 & 6.2 & 8.5   \\
    & Unsup & 3.2 & 8.6 & 4.6 & \bf{11.8} & 4.7 & 11.7 & 4.8 & 11.9 \\
    & Semi & 4.3 & 5.5 & \bf{6.8} & \bf{11.8} & 6.8 & 12.0 & 6.9 & 12.1 \\
    \hline
    \multirow{3}{*}{4-mic} & Sup & 0.5 & -7.3 & 4.8 & 6.1 & 6.3 & 9.7 & 6.3 & 9.7   \\
    & Unsup & 1.5 & 3.1 & 4.2 & 10.9 & 4.8 & 12.8 & 4.9 & 13.0 \\
    & Semi & 2.6 & 2.3 & 5.9 & 10.5 & \bf{7.0} & \bf{12.8} & 7.0 & 12.8 \\
    \hline
    \multirow{3}{*}{8-mic} & Sup & -0.8 & -8.6 & 3.8 & 4.4 & 5.8 & 9.9 & 6.3 & 10.6   \\
    & Unsup & 1.3 & 3.0 & 4.3 & 10.9 & 4.9 & 12.9 & 5.0 & \bf{13.6} \\
    & Semi & 2.5 & 3.4 & 5.6 & 9.5 & 6.6 & 11.4 & \bf{7.1} & 11.2 \\
    \thline
    \end{tabular}
    }
    \label{tab:crosseval}
\end{table}

\section{Results}
\label{sec:results}
\begin{table}
\small
\centering
\caption{AMI data results. ``S1'' and ``S2'' refer to SI-SNRi for the full-duration speaker and the overlapping speaker, respectively. The first microphone is used as the reference for reference-based metrics. For full synthetic AMI, the absolute input SI-SNRs are 0.5 dB for S1 and -9.2 dB for S2, which are used in the SI-SNRi computation. ``Warm'' indicates loading the model weight pre-trained with MixIT on 1600 hours of YFCC100M data (single-channel).  The pooled 95\% confidence intervals are $\pm 1.1$ for the MUSHRA and $\pm 2.2$ for the MUSHIRA ratings.}
\label{tab:ami}
\scalebox{0.78}{
\begin{tabular}{c|c|c|c|c|c|c} \hline
\multicolumn{3}{c|}{Model Configuration} & \multicolumn{3}{c|}{Synthetic AMI} & \multicolumn{1}{c}{Real AMI} \\
\multicolumn{1}{c}{Sup PIT} & \multicolumn{1}{c}{Unsup MixIT} & \multicolumn{1}{c|}{Warm} & \multicolumn{1}{c}{S1} & \multicolumn{1}{c}{S2} & \multicolumn{1}{c|}{MUSHRA} & \multicolumn{1}{c}{MUSHIRA} \\ \hline
\multicolumn{7}{c}{\bf{Baselines}} \\ \hline
\multicolumn{3}{c|}{Headset} & -- & -- & 96.6 & 93.4\\
\multicolumn{3}{c|}{Headset filtered to distant mic} & $\infty$ & $\infty$ & 64.5 & 50.0\\
\multicolumn{3}{c|}{Distant mic} & 0.0 & 0.0 & 33.1 & 38.8\\ 
\hline
\multicolumn{7}{c}{\bf{1-microphone}} \\ 
\hline
 -- & YFCC & --  & 2.1 & 2.5 & 29.1 & 29.8 \\
 -- & AMI & -- & 3.6 & 10.0 & 38.4 & \bf{43.5} \\
Mixed AMI & -- & -- & -1.0 & 6.7 & 39.4 & 38.9 \\
Synth AMI & -- & -- & 4.3 & 6.0 & 35.8 & 40.9 \\
Mixed AMI & AMI & -- & 0.0 & 10.2 & 41.8 & 39.8 \\
Synth AMI & AMI & -- & 6.0 & 10.3 & 39.0 & 41.8 \\
-- & AMI & YFCC & 3.7 & 9.8 & 40.4 & 41.1 \\
Mixed AMI & AMI & YFCC & -0.4 & 9.4 & 42.1 & 42.6 \\
Synth AMI & AMI & YFCC & \bf{6.4} & \bf{14.1} & \bf{42.9} & 41.7 \\
\hline
\multicolumn{7}{c}{\bf{4-microphone}} \\
\hline
-- & AMI & -- & 4.8 & 12.8 & 43.7 & 44.3 \\
Mixed AMI & -- & -- & 0.4 & 10.9 & 43.9 & 43.8 \\
Synth AMI & -- & -- & 6.3 & 9.7 & 37.5 & 38.2 \\
Mixed AMI & AMI & -- & 1.9 & 12.5 & 44.7 & 43.9 \\
Synth AMI & AMI & -- & 7.0 & 12.8 & 40.9 & \bf{46.2} \\
-- & AMI & YFCC & 4.5 & 12.0 & 44.2 & 44.3 \\
Mixed AMI & AMI & YFCC & 0.2 & 12.5 & 43.8 & 42.7 \\
Synth AMI & AMI & YFCC & \bf{7.2} & \bf{16.4} & \bf{46.5} & 46.1 \\
\hline \end{tabular}
}
\end{table}

Our TCN-TAC architecture allows models trained with any number of input microphones to be applied to data with a different number of microphones.  Table~\ref{tab:crosseval} cross-evaluates models trained on $N$ microphones on the evaluation sets for $M$ microphones, for $N, M \in [1, 2, 4, 8]$.   We observe that 1-mic trained models perform nearly the same no matter how many input mics are provided.  For $N > 1$, quality is best when the number of training microphones equals the number of input microphones, but models degrade gracefully when given a different number of microphones, and in some cases quality improves modestly when additional mic inputs are provided beyond what was used for training. We also observe that unsupervised learning outperforms supervised learning in most cases on speaker 2 (which we focus on because its reference is cleaner as described in Section \ref{sec:training}), and combining both training methods can further improve the performance.

Due to human evaluation capacity constraints, we were unable to do human listening eval of all models. Instead, in Table~\ref{tab:ami}, we take 1-microphone and 4-microphone models as examples and provide a comprehensive comparison of different model training configurations in terms of SI-SNRi scores on the fully synthetic AMI evaluation dataset, MUSHRA scores for a subset of a few hundred synthetic AMI examples, and MUSHIRA scores for about 100 real overlapping AMI examples.

In the 1-microphone subtable, unsupervised training with MixIT on AMI outperforms supervised training on either mixed AMI or synthetic AMI across most metrics. However, it falls short in terms of the MUSHRA score compared to supervised training on mixed AMI (38.4 v.s. 39.4), and in terms of SI-SNRi for speaker 1 compared to supervised training on synthetic AMI (3.6 v.s. 4.3). Combined supervised and unsupervised training further improves the SI-SNRi and MUSHRA scores on synthetic AMI, but does not improve the MUSHIRA score on real AMI. 

The model trained with MixIT on YFCC100M performs quite poorly on the AMI eval set across all metrics.    However, using the YFCC100M-trained model to warm-start improves MUSHRA and MUSHIRA scores when using mixed AMI as the supervised dataset, while improving SI-SNRi and MUSHRA scores when using synthetic AMI as the supervised dataset.

Each 1-microphone model has a 4-microphone counterpart, with the exception of the single-channel separation model trained on YFCC100M, which is evaluated alongside other models and also used to warm-start some other configurations. Each pair of 1-microphone and 4-microphone models share the same learning strategies, with the only difference being that the 4-microphone models take advantage of multi-channel signals by using TAC modules to exploit spatial features for better separation. (TAC modules are present in both 1- and 4-microphone models, but they only have the effect of transferring information across channels in the 4-microphone models.) Notice that all 4-microphone models significantly outperform their 1-microphone counterparts in terms of SI-SNRi, MUSHRA, and MUSHIRA scores, except for the 4-microphone model trained supervised on synthetic AMI, which had a lower MUSHIRA score (38.2) compared to 1-microphone (42.9).

Overall, the results demonstrate that multi-channel models achieve better separation performance in supervised learning, unsupervised learning, and their combination. This confirms that multi-channel models can take advantage of unsupervised learning to adapt on real-world multi-channel recordings. Notably, when training the multi-channel model using both MC-MixIT unsupervised and PIT supervised on synthetic AMI, warm-starting from a model pre-trained with MixIT on monaural YFCC100M achieves significant improvement across all synthetic AMI eval metrics. It is likely YFMCC100m contains rich acoustics including diverse speakers and environmental conditions. In the pre-training stage, the model focuses on separating sources using solely spectral-temporal information and ends with an effective model weight initialization which benefits the model further exploring spatial information to improve separation in the next stage. Warm-starting a 4-microphone model using a 1-microphone model is feasible thanks to the architecture's invariance to the number of microphones. This highlights the potential of pre-training models on a large amount of general audio data that contains a wide variety of real-world speech and then adapting these models on a smaller number of domain-specific speech recordings from multi-mic arrays.

Finally, the most effective configuration for achieving optimal performance is the multi-microphone model that is pre-trained on YFCC100M and then employs semi-supervised training with PIT on synthetic AMI and MixIT on real AMI.

Audio demos are provided at \url{https://google-research.github.io/sound-separation/papers/mcmixit}.

\section{Conclusions}
\label{sec:conclusions}

In this paper, we generalized a single-channel unsupervised learning method, MixIT, to multi-channel settings.  We also introduced a flexible multi-channel input and multi-channel output separation model applicable to arbitrary numbers of microphones and array geometries. We showed that multi-channel MixIT enables model adaptation on real-world multi-channel unlabeled spontaneous speech recordings. Leveraging the flexibility of our model to train on data with different numbers of channels, our best-performing system combines pre-training with MixIT on a large amount of single-channel data from YFCC100M, supervised training with PIT on synthetic multi-channel data, and unsupervised training with MixIT on multi-channel target domain data. In the future, we plan to explore how to create better supervised datasets that can enhance the separation ability without compromising the model's generalization to target domain data, and to investigate using larger and more diverse amounts of open-domain data to improve separation performance. 

\newpage

\newpage
\bibliographystyle{IEEEtran}
{\footnotesize\bibliography{refs}}

\end{document}